\documentclass[a4paper,fleqn,usenatbib]{mnras}

\usepackage[T1]{fontenc}
\usepackage{ae,aecompl}
\usepackage{graphicx}
\usepackage{amsmath}
\usepackage{amssymb}
\usepackage{cases}
\usepackage{xspace}

\newcommand{\Nbody}{$N$-body\xspace}

\newcommand{\secref}[1]{Section~\ref{#1}}
\newcommand{\figref}[1]{Fig.~\ref{#1}}

\newcommand{\equref}[1]{equation~\eqref{#1}}

\defcitealias{gieles2014}{G14}

\title[Cluster on eccentric orbit]{Evolution of star clusters on eccentric orbits: semi-analytical approach}

\author[Ebrahimi et al.]
{Hamid Ebrahimi$^{1}$\thanks{E-mail:  \mbox{h.ebrahimi@iasbs.ac.ir} (HE); \mbox{haghi@iasbs.ac.ir} (HH)},
Hosein Haghi$^{1}$, Pouria Khalaj$^{2}$, Akram Hasani Zonoozi$^{1}$,
\newauthor
Ghasem Safaei$^{1}$\\
$^{1}$Department of Physics, Institute for Advanced Studies in Basic Sciences (IASBS), PO Box 11365-9161, Zanjan, Iran\\
$^{2}$Universit\'e Grenoble Alpes, CNRS, IPAG, 38000 Grenoble, France \\}

\begin{document}

\date{Accepted .... Received}

\pagerange{\pageref{firstpage}--\pageref{lastpage}} \pubyear{2017}

\maketitle

\label{firstpage}

\maketitle

\begin{abstract}
We study the dynamical evolution of star clusters on eccentric orbits using a semi-analytical approach. In particular we adapt and extend the equations of \textsc{emacss} code, introduced by Gieles et al. (2014), to work with eccentric orbits. We follow the evolution of star clusters in terms of mass, half-mass radius, core radius, Jacobi radius and the total energy over their dissolution time. Moreover, we compare the results of our semi-analytical models against \Nbody computations of clusters with various initial half-mass radius, number of stars and orbital eccentricity to cover both tidally filling and under-filling systems. The evolution profiles of clusters obtained by our semi-analytical approach closely follow those of \Nbody simulations in different evolutionary phases of star clusters, from pre-collapse to post-collapse. Given that the average runtime of our semi-analytical models is significantly less than that of \Nbody models, our approach makes it feasible to study the evolution of large samples of globular clusters on eccentric orbits.
\end{abstract}

\begin{keywords}
galaxies: star clusters: general -- globular clusters: general -- methods: numerical
\end{keywords}

\section{Introduction}
The evolution of star clusters is driven by internal and external processes. Two-body relaxation, stellar evolution and binary formation are the main internal drivers and the galactic tidal field is the most significant external factor to the disruption of star clusters \citep{spitzer1987}. Due to the complexity of star clusters, numerical and computer codes are often used to study their evolution. Among all available methods, direct \Nbody codes such as \textsc{NBODY6/7}, which includes many details (e.g. the dynamical formation of higher order systems, stellar evolution of single and binary stars as well as a 3D external tidal field) generally produce the most accurate results \citep{aarseth1999, heggie2003}.

Although the runtime of \Nbody simulations have decreased by the advent of GRAPE computer series \citep{Makino2001} and  Graphics Processing Units (GPUs, \citealt{nitadori2012}) and the direct modeling of some globular clusters (GCs) is now possible (e.g. \citealt{zonoozi2011, zonoozi2014, zonoozi2017}), clusters with $N\gtrsim10^5$ stars are still computationally expensive to simulate especially in the presence of primordial binaries \citep{wang2015, wang2016}. The other notable methods for simulating star clusters that are based on the statistical modeling of GCs include different techniques like Fokker-Planck, Monte Carlo and gaseous model simulations \citep{Spurzem2005}. Monte Carlo codes \citep{henon1971a, henon1971b, giersz1998} such as \textsc{MOCCA} \citep{hypki2013}
have been shown to follow the outcome of \Nbody simulations closely, albeit they need to be calibrated using \Nbody simulations first \citep{giersz2008}. The main advantage of such Monte Carlo codes is that their runtime scales with the number of stars ($N$), hence they are faster than direct \Nbody simulations whose runtime scales as $\mathcal{O}(N^{2})$.

In terms of analytical attempts to study the evolution of star clusters, one can mention the pioneering  analytical work of H{\'e}non who used a Fokker-Planck equation for gravitational encounters, for tidally limited \citep{henon1961} and isolated clusters \citep{henon1965}. H{\'e}non showed that a single-mass star cluster in a tidal field loses its mass in a self-similar manner and an isolated cluster keeps its mass and expands.

\cite{alexander2012} introduced \textsc{emacss}, a fast C++ code, to simulate the evolution of single-mass star clusters on circular orbits affected by two-body relaxation and an external tidal field. In the first version of the code, \textsc{emacss} solved two coupled differential equations for number of stars and half-mass radius and predicted the evolution of these parameters accurately in comparison with \Nbody results. In the second version of \textsc{emacss} \citep[hereafter G14]{gieles2014}, the core evolution of a single-mass star cluster was added and the pre-collapse phase or the unbalanced evolution which is an important phase in the evolution of star clusters in a strong tidal field is included. In the third version \citep{alexander2014}, the effect of stellar evolution given an initial mass function was added to \textsc{emacss}.

\textsc{emacss} is useful to study the evolution of a large sample  of star cluster since it is much faster than both Monte Carlo and \Nbody simulations. However, in its current form it is only limited to circular orbits. The Milky Way GCs with determined orbital properties have eccentric orbits \citep{dinescu1999, casetti-dinescu2007, casetti-dinescu2013}. The effect of eccentric orbits over the life-time of star clusters is not negligible \citep{Kupper15}. The escape rate of stars from a star cluster depends on its galactocentric distance. While a cluster passes through its perigalacticon, the stars gain enough energy to go to outer orbits due to tidal heating and shocking and can leave the cluster \citep{webb2013, webb2014}. \cite{baumgardt2003} evolved a large set of \Nbody simulations for star clusters both on circular and eccentric orbits. They showed that the dissolution time of star clusters on eccentric orbits is approximately proportional to those on circular orbit with the radius equal to the apogalactic distances as $ T_{diss}(e)\sim T_{diss}(R_A, 0)(1-e)$, where, $e$ is the orbital eccentricity and $T_{diss}(R_A, 0)$ is the lifetime of a cluster moving on a circular orbit with a radius equal to the apogalactic radius of the eccentric orbit. However, recently, \cite{xu-cai2016} updated this relation using a series of N-body simulations and found that the dissolution time of a star cluster on an eccentric orbit is proportional to a circular orbit with the radii equal to the semi-major axis as  $ T_{diss}(e)\sim T_{diss}(a, 0)(1-e^2)(1-ce^2)$ where $c\simeq0.5$ and $T_{diss}(a, 0)$ is the lifetime of a cluster moving on a circular orbit with a radius equal to the semi-major axis of the eccentric orbits.

In the present paper, we aim to extend the work of \citet{alexander2014} to study star clusters moving on eccentric orbits. In particular we achieve this by adapting the equations of \textsc{emacss} to work with eccentric orbits. In order to check the accuracy of our method, we compare our results with \Nbody models.

The present paper is organized as follows. In \secref{Sec:2} we describe the equations that we have used to model the evolution of star clusters on eccentric orbits. We compare the results of our semi-analytical models against \Nbody simulations in \secref{Sec:3}. Finally, in \secref{Sec:4} we conclude and summarize our work.

\section{Overview of semi-analytical approach}\label{Sec:2}

In this section, we set out the theoretical framework that is used
to describe the evolution of the mass, half-mass radius, and core radius of star clusters. We start with introducing the key parameters and equations that determine the evolution of star clusters,  the time-scales and definitions that is used in this paper and overview the basic physical details of escape considered by our prescription.

\subsection{Equations of Evolution}\label{Sec:2.1}

We assume a single-mass star cluster with total mass $M$ including $N=M/m$ stars, where $m$ is the mass of each star and is constant for all stars. In the present paper we have not considered the effect of an initial mass function, stellar evolution or primordial binaries. This enable us to isolate the influence of eccentric orbits compared to circular orbits on the evolution of star clusters. We assume that star clusters move on eccentric orbits around a point-mass host galaxy with mass $M_G$, located at galactocentric distance $R_G(t)$ which changes as a function of time.

The total energy of a star cluster in virial equilibrium is given by \citep{spitzer1987}

\begin{equation} \label{total energy}
    E=-\kappa\dfrac{GM^{2}}{r_h},
\end{equation}
where $G$ is the gravitational constant, $r_h$ is the half-mass radius and $\kappa$ is the form factor
that depends on the density profile of the star cluster and can be written as $\kappa=r_h/(4r_v)$, where $r_v$ is the virial radius \citepalias{gieles2014}.

The theoretical starting point of this paper is based on the diffusion of energy in a star cluster. In the lack of energy sources such as stellar evolution and primordial binaries, the single-mass star cluster contracts in the early phase of its evolution. Therefore the inner region of cluster heats up and loses energy to the outer halo. This process continues to the extent that the core radius ($r_c$) becomes very small and the core density grows significantly \citep{takahashi1995}. This process is called core collapse and it stops when the first hard binary forms. The evolution phase before the core collapse is called pre-collapse phase or unbalanced evolution. After the core collapse, the energy flows from the core into the halo with a constant rate by the two-body relaxation \citep{alexander2012}. This evolution phase is called post-collapse or balanced evolution. The flux of energy through the star cluster changes its characteristic parameters, i.e. mass, half-mass radius, core radius and density profile. The time-scale for dynamical evolution of single-mass cluster is determined by half-mass relaxation time, $\tau_{rh}$, as \citep{spitzer1971}

\begin{equation} \label{relaxation time}
    \tau_{rh}=0.138\dfrac{N^{1/2}r_h^{3/2}}{\sqrt{Gm}\ln({\gamma N})},
\end{equation}
where $\ln({\gamma N})$ is the Coulomb logarithm with $\gamma\approx 0.11$ for single-mass star clusters
\citep{giersz1994}. Following \citetalias{gieles2014}, we can define five dimensionless parameters for the evolution of
each of the characteristic parameters of star clusters per $\tau_{rh}$ as

\begin{numcases}
    \
    \epsilon\equiv-(\dot{E}/E)\tau_{rh},  \label{energy production rate}
    \\
    \xi\equiv-(\dot{N}/N)\tau_{rh}=-(\dot{M}/M)\tau_{rh},  \label{escape rate}
    \\
    \mu\equiv(\dot{r}_h/r_h)\tau_{rh},  \label{half-mass radius rate}
    \\
    \lambda\equiv(\dot{\kappa}/\kappa)\tau_{rh}, \label{form factor rate}
    \\
    \delta\equiv(\dot{r}_c/r_c)\tau_{rh}.  \label{core radius rate}
    \
\end{numcases}
Note that the equality in \equref{escape rate} is the direct result of the constant mean mass of stars.
In \equref{energy production rate}, $\epsilon$ represents the energy production rate. Mass loss from
the star cluster can be obtained from $\xi$ in \equref{escape rate} which remains positive during the life-time of the shrinking star cluster. According to equations (\ref{half-mass radius rate}) and (\ref{core radius rate}), evolution of
the half-mass radius and the core radius ($r_c$) can be evaluated from $\mu$ and $\delta$, respectively.
Taking the time derivative on both sides of \equref{total energy} and using equations
(\ref{energy production rate}) to (\ref{form factor rate}), one can relate four of the dimensionless
parameters as

\begin{equation} \label{rates relation}
    \epsilon=-\lambda+\mu+2\xi.
\end{equation}
Note that $\delta$ does not appear in \equref{rates relation}. This is due to the fact that the total energy does not depend on the core radius directly (see \equref{total energy}). The evolution of the core radius on eccentric orbit is one of the goals of this paper and we can evaluate it from the equations described in section \ref{Sec:2.3}.

The apogalactic ($R_A$) and perigalactic ($R_P$) radii of the star cluster orbit are related to eccentricity $e$, and semi-major axis $a$, by $R_A=a(1+e)$ and
$R_P=a(1-e)$, respectively. The galactocentric distance of the star cluster and elapsed time are related via the following equations \citep{binney2008}

\begin{numcases}
    \
    R_G=a(1-e\cos\eta),  \label{galactocentric distance}
    \\
    t=(T/2\pi)[(\eta -\eta_0)-e\sin(\eta -\eta_0)],  \label{kepler equation}
    \
\end{numcases}
where $\eta$ is the eccentric anomaly. Equation  (\ref{kepler equation}) is the Kepler's equation. In it  $T=2\pi\sqrt{a^{3}/(GM_G)}$ is the orbital period and $\eta_0$ is the initial phase, i.e. if $\eta_0=0$, the star cluster begins its evolution at perigalacticon. We use the Newton-Raphson  method to numerically solve the Kepler's equation and determine $R_G(t)$ as a function of time.

The external tidal field puts a limiting boundary for the star cluster so that if stars gain enough energy to pass this boundary, they become unbound from the star cluster and scatter into the host galaxy. This boundary is specified by the Jacobi radius $r_J$.
For a star cluster on an eccentric orbit around a point-mass galaxy, the instantaneous Jacobi radius is given by \citet{ernst2013}

\begin{equation}  \label{jacobi radius}
    r_J= \bigg(\dfrac{M}{M_G}\bigg)^{1/3} \bigg(\dfrac{aR_G^{4}}{R_AR_P+2aR_G}\bigg)^{1/3}.
\end{equation}
In this equation $R_G$ is time-dependent and can be calculated using \equref{galactocentric distance} and \equref{kepler equation}.

It should be noted that a strict theoretical definition of the Jacobi radius exists only for circular orbits and for a cluster on an eccentric orbit there is no clear  definition of the Jacobi radius. As a result, \equref{jacobi radius} is not accurate for the case of the eccentric orbits (See Sec. \ref{Sec:2.2} for more discussion).

\begin{table}
    \centering
    \caption{Initial parameters and inputs of each performed semi-analytical models and N-body runs. The evolution of
each model is shown in the figures which is identified in the first column. Columns 2 and 3 are the initial number of stars and initial half-mass radius, respectively. The orbital parameters are given in columns 4 to 6. The last two columns are the initial Jacobi radius and filling factor that are analytically calculated. See the text for more details. }
    \label{params}
    \begin{tabular}{cccccccccc}
        \hline
        Figures &$N_0$ &$r_{h0}$ &$R_P$ & $R_A$ & $e$ & $r_{J0}$ & $\mathcal{R}_{hJ0}$ \\
        \# &  & [pc] & [kpc] & [kpc] &  &  [pc] &    \\
        \hline
        \ref{evolution1} & 5\,000 & 3 & 10 & 30 & 0.5 & 19.23 & 0.16 \\
        \ref{evolution1} & 8\,000 & 3 & 10 & 30 & 0.5 & 22.50 & 0.13 \\
        \ref{evolution1} & 16\,000 & 3 & 10 & 30 & 0.5 & 28.36 & 0.11 \\
        \hline
        \ref{evolution3} & 4\,000 & 0.3 & 10 & 30 & 0.5 & 17.87 & 0.02 \\
        \ref{evolution3} & 4\,000 & 1 & 10 & 30 & 0.5 & 17.87 & 0.03  \\
        \ref{evolution3} & 4\,000 & 3 & 10 & 30 & 0.5 & 17.87 & 0.17 \\
        \ref{evolution3} & 4\,000 & 6 & 10 & 30 & 0.5 & 17.87 & 0.34 \\
        \hline
        \ref{diff-ecc1} & 8\,000 & 3 & 30 & 30 & 0 & 71.13 & 0.04 \\
        \ref{diff-ecc1} & 8\,000 & 3 & 16.1 & 30 & 0.3 & 37.10 & 0.08  \\
        \ref{diff-ecc1} & 8\,000 & 3 & 5.3 & 30 & 0.7 & 11.66 & 0.26 \\
        \hline
        \ref{diff-ecc3} & 8\,000 & 3 & 15 & 15 & 0 & 35.57 & 0.08 \\
        \ref{diff-ecc3} & 8\,000 & 3 & 8.1 & 15 & 0.3 & 18.60 & 0.16  \\
        \ref{diff-ecc3} & 8\,000 & 3 & 5 & 15 & 0.5 & 11.20 & 0.26 \\
        \ref{diff-ecc3} & 8\,000 & 3 & 2.65 & 15 & 0.7 & 5.86 & 0.51\\
        \hline

    \end{tabular}
\end{table}

\subsection{Review of the required formulas and initial conditions}\label{Sec:2.3}

To determine the time-evolution of the characteristic parameters of star clusters, we solve the five coupled  differential equations (\ref{energy production rate}) to (\ref{core radius rate}). To achieve this we need additional expressions to relate the five dimensionless parameters ($\epsilon, \xi, \mu, \lambda, \delta$) to other parameters of the star clusters. Here, we review the additional formulas needed to relate the five coupled differential equations (\ref{energy production rate}) to (\ref{core radius rate}) as given by \citetalias{gieles2014}. For clarity we follow the notation of \citetalias{gieles2014}.

First of all, It is useful to define three ratios as $\mathcal{R}_{hJ}\equiv r_h/r_J$, $\mathcal{R}_{vJ}\equiv r_v/r_J$, $\mathcal{R}_{ch}\equiv r_c/r_h$. The ratio $\mathcal{R}_{hJ}$ (filling factor) denotes the strength of the galactic tidal field, i.e. the star clusters with initial $\mathcal{R}_{hJ}\gtrsim0.1$ are initially Roche volume filling clusters and those with initial $\mathcal{R}_{hJ}<0.1$ are initially Roche volume under-filling clusters \citep{alexander2013}. However following \citetalias{gieles2014}, we use $\mathcal{R}_{vJ}$ instead of $\mathcal{R}_{hJ}$ in mass-loss rate (\equref{P}).

We differentiate between the pre-collapse phase (unbalanced evolution, hereafter UE) and the post-collapse phase (balanced evolution, hereafter BE). Therefore, we present two sets of equations for each of the five dimensionless parameters ($\epsilon, \xi, \mu, \lambda, \delta$). The transition from UE to BE occurs at the core collapse time. The core collapse occurs when
$\mathcal{R}_{ch}$ in balanced phase reaches the value that is given by \citep[G14]{heggie2003}

\begin{equation} \label{core collapse time}
    \mathcal{R}_{ch}=\bigg(\dfrac{N_2}{N}+\dfrac{N_2}{N_3}\bigg)^{2/3}
\end{equation}
where $N_2$ and $N_3$ are two free parameters and are chosen to be $N_2=12$ and $N_3=15\,000$. During the pre-collapse phase, the core radius decreases and the core density, $\rho_c$, grows. In the post-collapse phase, the core radius
increases and the core density reduces as $r_c^{-2}$. The relation of the core density can be written as

\begin{numcases}{\rho_c =}
    \rho_{c0} r_c^{-\alpha} & for UE;  \label{unbalanced core density}
    \\
    \rho_h \mathcal{R}_{ch}^{-2} & for BE.  \label{balanced core density}
\end{numcases}
In \equref{unbalanced core density}, the relation between the core radius and density is a power-law relation whose exponent $\alpha$, has been determined in several studies, e.g. $\alpha=2.21$ \citep{heggie1988}, $\alpha=2.23$ \citep{takahashi1995}, $\alpha=2.26$ \citep{baumgardt2003a}
and $\alpha=2.20$ \citepalias{gieles2014}. We adopt $\alpha = 2.2$ and $\rho_{c0}=0.055 M_0 r_{v0}^{-0.8} $.  In \equref{balanced core density}, $\rho_h$ is the half-mass density and is defined as $\rho_h\equiv3M/(8\pi r_h^{3})$.

The energy production rate, $\epsilon$, in the UE is time-dependent and increases as stars escape. In the BE this rate is approximately constant or based on \cite{alexander2012} notation $\zeta\simeq0.1$.
Thus we can express the energy production rate as

\begin{numcases}{}
    \epsilon=(\mathcal{R}_{hJ}/\kappa)\xi & for UE;  \label{unbalanced energy rate}
    \\
    \epsilon=\zeta \simeq 0.1 & for BE.  \label{balanced energy rate}
\end{numcases}
In \equref{unbalanced energy rate}, $\xi$ is the escape rate of stars from the star cluster which can be obtained as \citepalias{gieles2014}

\begin{equation} \label{escape rate 2}
    \xi=\mathcal{F}\xi_1(1-\mathcal{P})+(3/5)\zeta[ (f+(1-f)\mathcal{F}]\mathcal{P}.
\end{equation}
The second term on the right hand side of \equref{escape rate 2} accounts for the escape rate in a tidal field and the first term
denotes the mass-loss of isolated cluster with $\xi_1=0.0142$. In \equref{escape rate 2}, $f=0.3$ is the ratio of escape rate of a Roche volume filling cluster to that in BE, and $\mathcal{F}$  causes a smooth transition from UE to BE and is given by

\begin{numcases}{\mathcal{F}=}
    \mathcal{R}_{ch}^{min}/\mathcal{R}_{ch} & for UE;  \label{F}
    \\
    1 & for BE.  \label{F}
\end{numcases}
where $\mathcal{R}_{ch}^{min}$  is the minimum value of $\mathcal{R}_{ch}$ that occurs at core collapse time and is obtained by \equref{core collapse time}.

Due to the fact that the mechanism of escape rate changes in the case of eccentric orbit, we need to alter the $\mathcal{P}$-parameter in \equref{escape rate 2}. \citet{alexander2012} defined this parameter for circular orbit as

\begin{equation} \label{P}
    \mathcal{P}(e=0)=\bigg(\dfrac{\mathcal{R}_{vJ}}{\mathcal{R}_{vJ1}}\bigg)^{z}\bigg[\dfrac{N\ln (\gamma N_1)}{N_1\ln (\gamma N)}\bigg]^{1-x}.
\end{equation}
The parenthesis part in \equref{P} is the result of integration over the Maxwellian velocity distribution to obtain the fraction of stars with enough escape energy \citep{gieles2008} and the bracket part is proportional to the time delay of the stars which has gained enough energy to escape from the star cluster \citep{baumgardt2001}. The values of the free parameters in \equref{P} are chosen by \citetalias{gieles2014} to be $z=1.61$, $x=0.75$, $\mathcal{R}_{vJ1}=0.145$.  $N_1$ is the scaling value of $N$ that is  tuned by \citetalias{gieles2014} to be  $N_1= 15\,000$  to match the N-body simulations for circular orbits.  The values of $N_1$, $\mathcal{R}_{vJ1}$, and consequently $\mathcal{P}$  for the eccentric orbits are different from those in \citetalias{gieles2014} for circular orbit. In section \ref{Sec:2.4} we introduce a new method for fixing the $\mathcal{P}-$parameter as a function of eccentricity (based on the findings by \cite{baumgardt2003} and \cite{xu-cai2016}  such that the semi-analytical models,  can reproduce the N-body simulation for different eccentricities.

For the evolution of the core radius we have

\begin{numcases}{\delta=}
    \delta_1+\delta_2\dfrac{\tau_{rh}}{\tau_{rc}} & for UE;  \label{unbalanced core radius rate}
    \\
    \dfrac{2}{3}\xi\bigg(1+\dfrac{N}{N_3}\bigg)^{-1}+\mu & for BE.  \label{balanced core radius rate}
\end{numcases}
In \equref{unbalanced core radius rate}, $\delta_1=-0.09$, $\delta_2=-0.002$ and $\tau_{rc}$ is the core relaxation time and is defined as \citep{spitzer1971}

\begin{equation} \label{core relaxation time}
    \tau_{rc}=\dfrac{\sigma_c^{3}}{15.4G^{2}m\rho_c \ln(\gamma N)},
\end{equation}
where $\sigma_c^{2}=(8/3)\pi G\rho_c r_c^{2}$. Note that we can derive \equref{balanced core radius rate} by combining equations (\ref{half-mass radius rate}), (\ref{core radius rate}) and (\ref{core collapse time}).

Letting $\mathcal{K}\equiv \mathrm{d}\ln \kappa / \mathrm{d} \ln \mathcal{R}_{ch}$, the evolution of half-mass radius can be described as \citepalias{gieles2014}

\begin{numcases}{\mu=}
    \dfrac{(\mathcal{R}_{hJ}/\kappa -2)\xi + \mathcal{K}\delta}{1+\mathcal{K}} & for UE;  \label{unbalanced half-mass radius rate}
    \\
    \zeta +\bigg[\dfrac{2}{3}\mathcal{K}\bigg(1+\dfrac{N}{N_3}\bigg)^{-1} -2\bigg]\xi & for BE.  \label{balanced half-mass radius rate}
\end{numcases}
The combination of equations (\ref{rates relation}) and (\ref{unbalanced energy rate}) leads to \equref{unbalanced half-mass radius rate}. Moreover, one can derive \equref{balanced half-mass radius rate} using equations (\ref{rates relation}), (\ref{balanced energy rate}) and (\ref{balanced core radius rate}).

Finally, we should determine the evolution of form factor. \citetalias{gieles2014} proposed a relation between form factor and $\mathcal{R}_{ch}$ by
fitting \Nbody results. They showed that the form factor in this relation is not necessarily equal to $\kappa$. Labeled as $\kappa(\mathcal{R}_{ch})$, the fitting function is

\begin{equation} \label{Re_ch(k)}
    \kappa(\mathcal{R}_{ch})=\kappa_1+(\kappa_0-\kappa_1) \,\mathrm{erf}(\mathcal{R}_{ch}/\mathcal{R}_{ch0}).
\end{equation}
Here, the three free parameters have different values in each phase: $\kappa_0=r_{h0}/(4r_{v0})$, $\kappa_1=0.295$ and $\mathcal{R}_{ch0}=0.100$
for UE and  $\kappa_0=0.200$, $\kappa_1=0.265$ and $\mathcal{R}_{ch0}=0.220$ for BE.

Using the logarithmic derivative of $\mathcal{K}$ and \equref{Re_ch(k)}, we have
\begin{equation} \label{K}
    \mathcal{K}=\dfrac{2(\kappa_0-\kappa_1)\mathcal{R}_{ch} \exp(-\mathcal{R}_{ch}^{2}/\mathcal{R}_{ch0}^{2})}{\sqrt{\pi}\kappa\mathcal{R}_{ch0}}.
\end{equation}
Using the definition of $\mathcal{K}$ and equations (\ref{half-mass radius rate}) to (\ref{core radius rate}), we can derive the evolution rate of the form factor as

\begin{numcases}{\lambda=}
    \mathcal{K}(\delta -\mu) & for UE;  \label{unbalanced form factor rate}
    \\
    \mathcal{K}(\delta -\mu)+\dfrac{\kappa(\mathcal{R}_{ch})- \kappa}{\kappa(\mathcal{R}_{ch})} & for BE.  \label{balanced form factor rate}
\end{numcases}

We have written a code which numerically solves the coupled differential equations for an eccentric orbit{\footnote{The code can be obtained by contacting the authors.}} using fourth-order Runge-Kutta with a constant time step of $\Delta t=1$ Myr.
For each model the input parameters are $N_0$ (initial number of stars), $r_{h0}$ (initial half-mass radius), $\eta_0$ (initial phase), $R_A$ and $R_P$. The initial
values of mass and form factor can be given as $M_0=mN_0$ and $\kappa_0=r_{h0}/(4r_{v0})$, respectively. The initial values of core and virial radius for a Plummer density profile are $r_{c0}\simeq0.4r_{h0}$ and $r_{v0}\simeq1.3r_{h0}$, respectively \citep[Table 8.1]{heggie2003}.

\begin{figure*}
\begin{center}
\includegraphics[width=165mm]{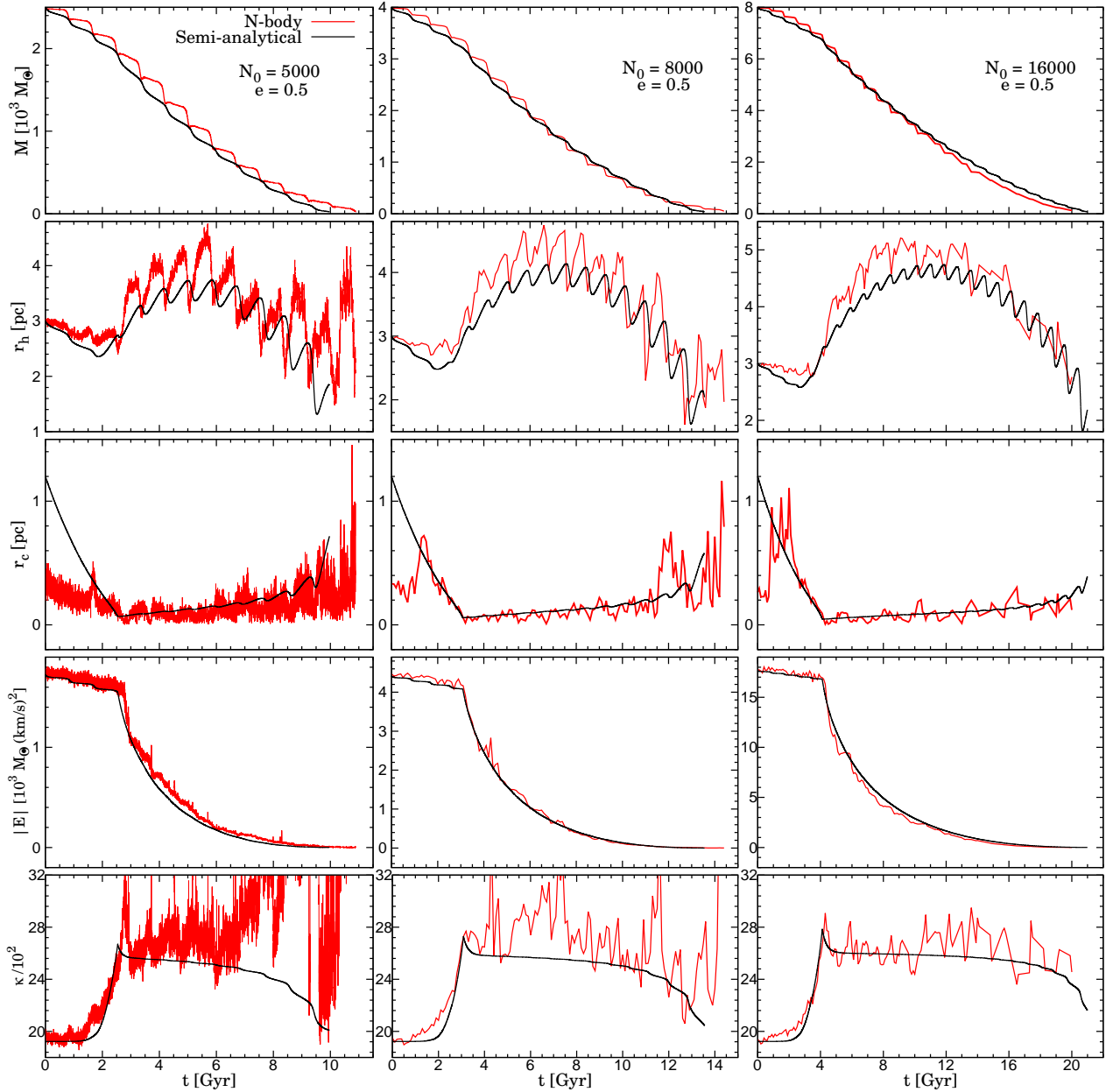}
\caption{Semi-analytical models vs. \Nbody simulation for the evolution of mass, half-mass radius, core radius, total energy, and form factor. In all models, the evolutions start at perigalacticon and we assumed the initial half-mass radius, eccentricity, apogalactic and perigalactic distance have the same values as $r_{h0}=3$\,pc, $e=0.5$, $R_A=30$\,kpc and $R_P=10$\,kpc, respectively. The columns represent the star cluster evolution for $N_0=5000$, $8000$ and $16000$ single-mass stars. }
\label{evolution1}
\end{center}
\end{figure*}

\begin{figure*}
\begin{center}
\includegraphics[width=165mm]{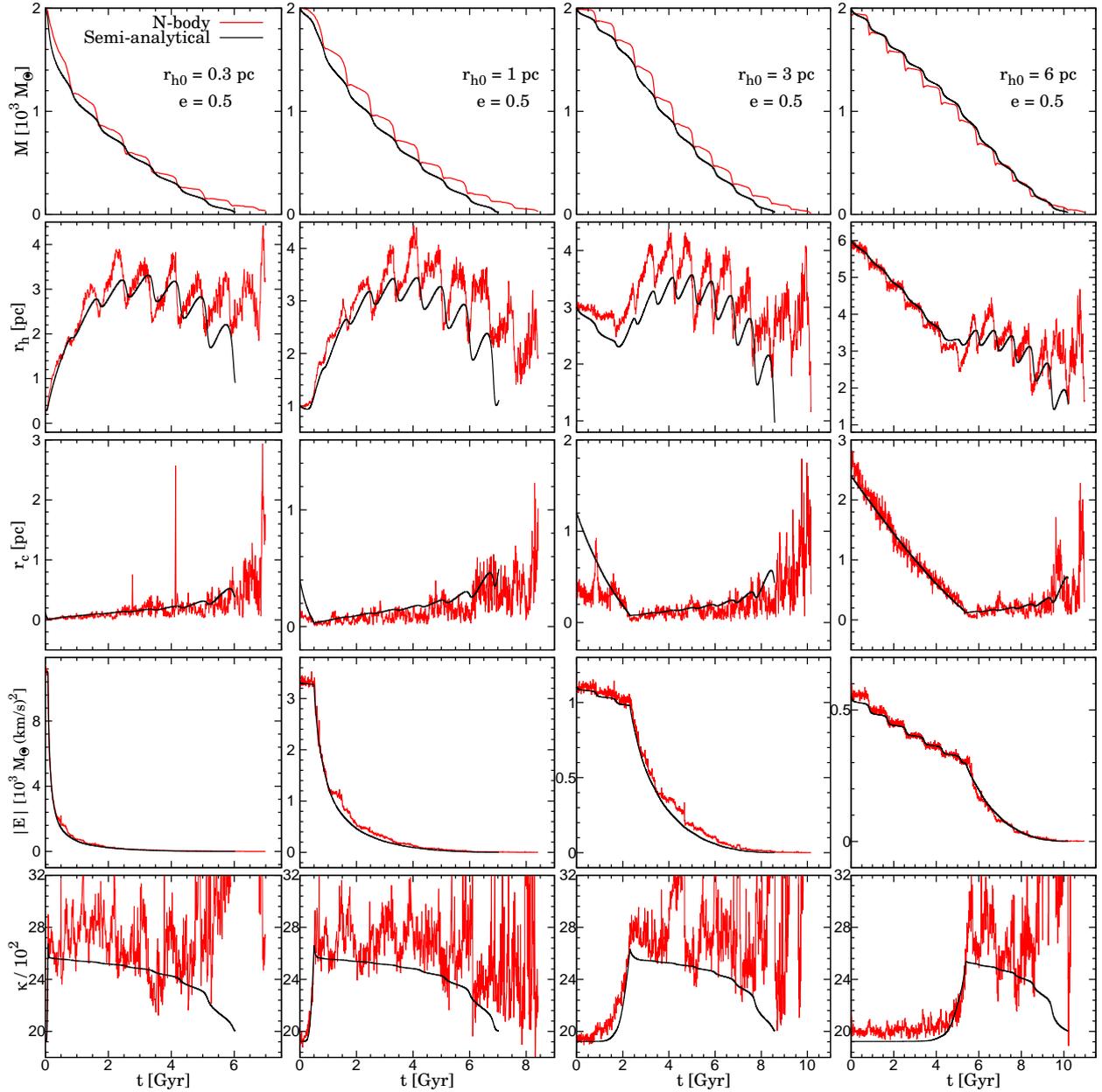}
\caption{Same as \figref{evolution1} but we assumed the initial number of stars, orbital eccentricity, apogalactic and perigalactic distance have the same values as $N_0=4\,000$, $e=0.5$,  $R_A=30$\,kpc
and $R_P=10$\,kpc, respectively. The columns represents the star cluster evolution with $r_{h0}=0.3$\,pc, $r_{h0}=1$\,pc and $r_{h0}=3$\,pc and $r_{h0}=6$\,pc.}
\label{evolution3}
\end{center}
\end{figure*}

\subsection{Determination of $\mathcal{P}-$parameter for eccentric orbits}\label{Sec:2.4}

Apart from the initial conditions,  the lifetime of the cluster is proportional to ${\mathcal{R}^z_{vJ1}} [N_1/\ln(\gamma N_1)]^{1-x}$ (see e.g., equation 23 in \citealt{alexander2012} or a simplified lifetime equation A10 in \citealt{gieles2011}). According to \citealt{xu-cai2016}, there is a relation between the lifetime of a star cluster  ($T_{diss}(a,e)$) on eccentric orbit with half major axis $a$, and the lifetime of a star cluster on circular orbit with radius $a$, $T_{diss}(a,0)$, as

\begin{equation} \label{dissolution time}
    T_{diss}(a,e)=T_{diss}(a,0)(1-e^2)(1-ce^2),
\end{equation}
where  $e$ is the eccentricity and $c\simeq0.5$.

Therefore, one can obtain a formula for $\mathcal{P}$-parameter for a cluster on an eccentric orbit as a function of eccentricity  and the corresponding parameter for a cluster evolving on a circular orbit ($\mathcal{P}(0)$) by multiplying $\mathcal{P}(0)$ by $\mathcal{Q}(e)$.

\begin{equation} \label{P ref}
    \mathcal{P}(e)=\bigg(\dfrac{\mathcal{R}_{vJ}}{\mathcal{R}_{vJ1}}\bigg)^{z}\bigg[\dfrac{N\ln (\gamma N_1)}{N_1\ln (\gamma N)}\bigg]^{1-x} \mathcal{Q}(e).
\end{equation}
where, $\mathcal{Q}(e)$ is a function of eccentricity as

\begin{equation} \label{Q ref}
    \mathcal{Q}(e)=[(1-e^2)(1-ce^2)]^{-1}.
\end{equation}

Using the above formulae for all models with different initial masses, radii, and eccentricities, one can see  the shape of the $M(t)$, $r_h(t)$, and $r_c(t)$ curves of semi-analytical models  are in general agreement with the N-body simulation (see Figs. \ref{evolution1} to \ref{diff-ecc3}). Our results also indicate that the scaling of the star cluster's lifetime as a function of eccentricity as obtained by \cite{xu-cai2016} is very successful to estimate the lifetime of star clusters on eccentric orbits and to reproduce the N-body simulation for different eccentricities.\\

\section{Comparison with \Nbody models}\label{Sec:3}
In this section the evolution of star clusters resulted from our semi-analytical models are compared against those of N-body simulations.
We used the GPU-enabled version of the collisional fourth-order Hermite \Nbody code \textsc{nbody6} \citep{nitadori2012}. We chose a Plummer model \citep{plummer1911} in virial equilibrium as the initial density profile for our models. All stars have equal mass $m=0.5 M_\odot$. The stars with distances more than $2r_J$ were assumed to be unbound from the cluster and the simulations were terminated when less than one percent of stars remained in the clusters. The modeled star clusters are on eccentric orbits around a point-mass galaxy with mass $M_G=10^{11}M_{\bigodot}$. All sets of models are summarized in Table 1.

\subsection{Models with $e=0.5$ and different initial masses and half-mass radii}\label{Sec:3.1}

First, we performed three simulations with $N_0=$ 5000, 8000, 16000 and the same half-mass radius $r_{h0}=3$ pc. Moreover, four additional simulations are performed with different half-mass radii of $r_{h0}=0.3, 1, 3, 6$ pc and the same number of stars $N_0=$ 4000.  We assume that the apogalactic and perigalactic distances of the orbits are $R_A=30$\,kpc and $R_P=10$\,kpc and the evolution starts at the perigalacticon. Thus the eccentricity of all orbits is $e=0.5$ which corresponds to the mean value for the globular star clusters in the Milky Way \citep{dinescu1999, casetti-dinescu2007, casetti-dinescu2013}.

We first model three star clusters with different $N_0$ and equal half-mass radii $r_{h0}=3$\,pc. The evolution of mass, half-mass radius, core radius, total energy, and form factor are shown in \figref{evolution1}. As can be seen, the mass of these models decreases with a staircase pattern as expected from \citet{xu-cai2016}. This pattern is a result of the variation of the galactocentric distance with time which consequently causes the Jacobi radius to oscillate.  In other words, when passing through the perigalacticon (apogalacticon), the star cluster experiences the strongest (weakest) tidal field on its orbit and its Jacobi radius reaches its minimum (maximum) value. As a result, the mass loss rate is larger at perigalacticon than that at the apogalacticon and the staircase shape is built. The evolution of mass which is obtained by our semi-analytical model is in good agreement with \Nbody results.

According to the evolution of $r_h$, three evolutionary phases are recognized: (i) in the pre-collapse phase, because of the absence of energy source in core, $r_h$ decreases and the amplitude of the oscillation is lower. (ii) in the post collapse phase, the flux of energy from the core to the outer regions of the cluster (as a result of two-body relaxation) leads to the expansion of cluster and $r_h$ increases (expansion-dominated regime). (iii) if the star cluster resides in a tidal field, its expansion is eventually halted at some points and the evaporation of stars from cluster begins until complete dissolution (evaporation-dominated regime in post-collapse phase).

Although the oscillation amplitude of our models is lower than that of \Nbody simulations, in both approaches the amplitude of the oscillations of $r_h$ in the post-collapse phase is higher than those in the pre-collapse phase. Moreover, the amplitude of the oscillations is even higher in the latter part of evolution.

According to \figref{evolution1}, core radius of star clusters at core collapse moment is smaller than that at $t=0$ but it increases in the post-collapse phase. The core radius oscillates in \Nbody simulations. Our semi-analytical models do not reproduce these oscillations, since the oscillations come from the escape rate and the core evolution rate ($\delta$) does not depend on it (see \equref{unbalanced core radius rate}). The evolution of $r_c$ is in good agreement with \Nbody simulations in the post-collapse phase.

The total energy increases in both pre- and post-collapse phases but with different rates and the oscillation in energy occurs only in the pre-collapse phase in our semi-analytical models (\equref{unbalanced energy rate}). The form factor increases in the pre-collapse phase and then reaches a constant value in the post-collapse phase and at the end of the cluster life time decreases. The amplitude of the Jacobi radius decreases in all models which is related to the mass-loss rate and remarkably follow the \Nbody results.

Fig. \ref{evolution3}  shows the evolution of mass, half-mass radius, core radius, total energy, and form factor of four star clusters with different $r_{h0}$ and the same number of stars ($N_0=4\,000$). These figures show how our semi-analytical method works in different tidal regimes. From left to right, we model four star cluster with different initial half-mass radii. For these models the value of $\mathcal{R}_{hJ}=r_h/r_J$ increases from 0.015 to 0.30 covering both weak and strong tidal fields. According to the classification of star clusters in a tidal field in \cite{alexander2013}, the first two models are Roche volume under-filling clusters and the other models are Roche volume filling clusters. The evolution of mass in these models show that the mass-loss rate in the early stages of evolution of Roche volume filling clusters in both \Nbody simulations and our semi-analytical models are higher. For instance, in the very weak tidal field, cluster loses around $~ 40\%$ of its mass during the first orbital period. The Roche-volume under-filling clusters reach core collapse after just a few Myrs, whereas the Roche-volume filling clusters spend a longer time in the pre-collapse phase (few orbital periods). In the strong tidal field, the star cluster spends roughly half of its lifetime in the pre-collapse phase. The dissolution time of these models increase with $r_{h0}$. The form factor ranges between 0.19 and 0.26 approximately in all models.  In the Roche volume filling clusters the form factor remains nearly constant during the pre-collapse phase. This is because the core and half-mass radii decrease at the same rate and at the instant of the core collapse, $\kappa$ rises.

\begin{figure*}
\begin{center}
\includegraphics[width=165mm]{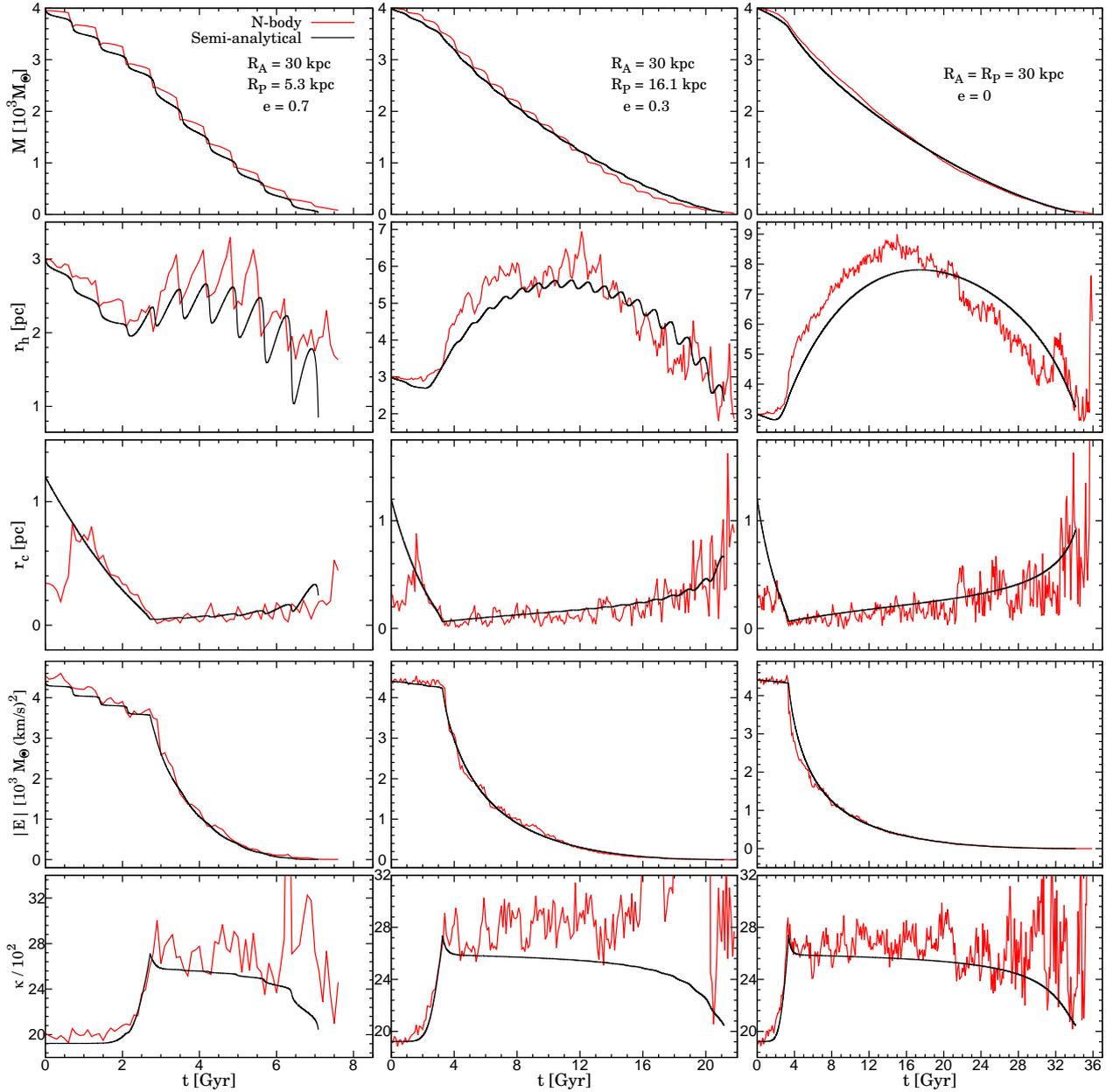}

\caption{Same as \figref{evolution1} but for star clusters with different orbital eccentricities of $e=0.7$, $e=0.3$ and $e=0$. In all plots, the evolution start at perigalacticon and we assumed the initial number of stars, half-mass radius and apogalactic distance have the same values as $N_0=8\,000$, $r_{h0}=3$\, pc and $R_A=30$\,kpc, respectively. }
\label{diff-ecc1}
\end{center}
\end{figure*}

\begin{figure*}
\begin{center}
\includegraphics[width=165mm]{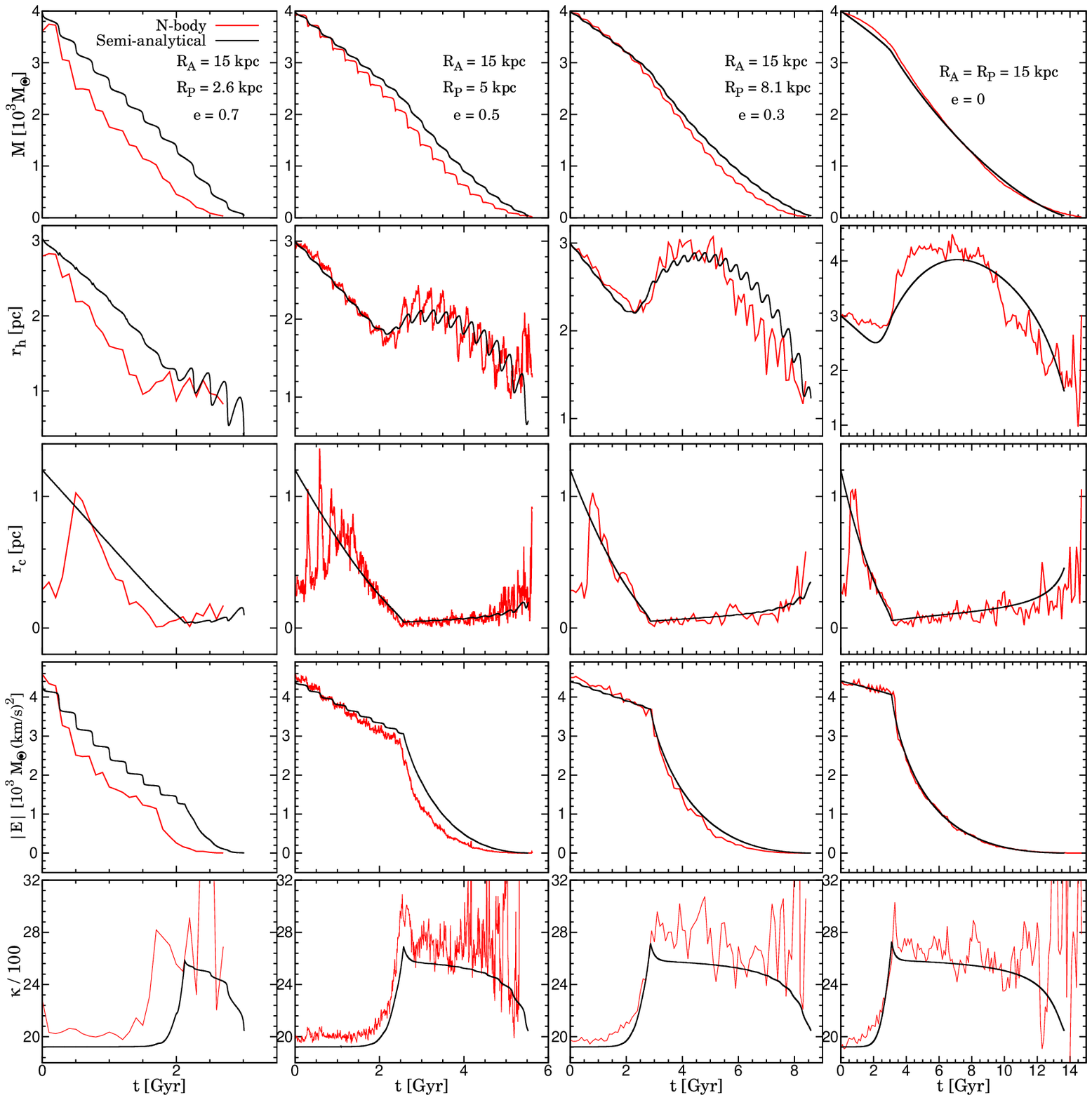}

\caption{Same as \figref{diff-ecc1} but for smaller apogalactic distance of $R_A=15$\,kpc.   The columns represent the star cluster evolution with $e=0.7$, $e=0.5$, $e=0.3$ and $e=0$, from left to right, respectively.}
\label{diff-ecc3}
\end{center}
\end{figure*}

\subsection{Models with different eccentricities}

In section \ref{Sec:3.1} all models are evolving on eccentric orbit with the specific eccentricity, $e=0.5$, and corresponding value of $N_1=2\,025$  was fixed for this eccentricity to match with the N-body simulation. In order to show that our semi-analytical model can reproduce the N-body simulation for different eccentricities we performed a new set of simulations with eccentricity $e= 0, 0.3, 0.5, 0.7$ (where eccentricity is defined as
$e=(R_A-R_P)/(R_A+R_P)$ and two different value of apocentric distance of $R_A=$ 15 and 30 kpc to cover both tidally filling and under-filling systems.

The results of our simulations for clusters with orbital eccentricities of $e=$0, 0.3, 0.7  and the same apogalactic distances of $R_A=30$ kpc are illustrated in Fig. \ref{diff-ecc1}. We also repeat the calculations for a lower value of apocentric distance of $R_A=15$ kpc to see how the higher filling factor affects the results (Fig. \ref{diff-ecc3}).

During the  perigalactic passage, the star clusters with high eccentricities experience a stronger tidal field which leads to an increase in their escape rate. Moreover, the  oscillation amplitude of the Jacobi radii is larger for star clusters with higher eccentricities. Thus the slope of decreasing mass is higher in high eccentricity cases and the staircase pattern of mass evolution is more clear. We  conclude that the lifetime of the star clusters increases with decreasing the eccentricity which is in agreement with the scaling formula for lifetime suggested by \cite{xu-cai2016}.

Comparing two models with the same eccentricity of $e=0.7$ but different pericentric and apocentric distances which corresponds to different filling factor, one can see easily that the more tidally filling cluster (Table \ref{params}) can not reproduce the N-body simulation. Therefore, our semi-analytical formalism is valid for both tidally filling and under-filling systems and violates only for the very high-eccentricity filling regime. A cluster with high eccentricity ($e=0.7$) spends less than half of its lifetime in pre-collapse phase but for the other two cases with the lower eccentricity of 0.5 and 0.3, the post-collapse phase dominates the pre-collapse phase and the half-mass radius grows up more in the first half part of post-collapse phase and consequently leads to a longer lifetime. \\

\section{Caveats: Tidal and limiting radii}\label{Sec:2.2}

As stated before tidal field is no longer static for a cluster with a non-circular orbit. This makes estimating the Jacobi radius for a GC on an eccentric orbit to be challenging. The variation of tidal field on a cluster moving on an eccentric orbit can be rapid at pericentre and results in a change in the shape of the Jacobi surface from a static approximation \citep{renaud2011} as well as an injection of additional energy into the star cluster \citep{weinberg1994a, weinberg1994b, weinberg1994c, kupper2010, kupper2012}. We therefore  model eccentric orbits in an approximate manner.

One approach in this regard is assuming that for a GC on an eccentric orbit, its tidal radius is imposed at perigalacticon ($R_p$) where the tidal field of the host galaxy is the strongest and the Jacobi radius has a minimum. That is,  derivative of $r_J$ is zero at this radius  \citep{von-hoerner1957, king1962}. This assumption is based on the fact that the internal relaxation time of the cluster is  greater than its orbital period for almost all observed GCs. Therefore, after stars outside the tidal radius at perigalacticon escape, the cluster would not be able to relax and expand before it returns to perigalacticon. In other words, the satellites are truncated during pericentre passages to the size indicated by the pericentric
tidal radius \citep{king1962,innanen1983}. However, this is indeed true for collisionless systems such as dwarf satellite galaxies that are orbiting around the host giant galaxy,  N-body simulations of collisional systems show that after the pericentre passage the satellite expands again and hence,  the \cite{king1962} conjecture that the satellites are trimmed at the pericentre and then remain unchanged is not valid \citep{gajda2016}. If the cluster moves on an eccentric orbit, stars outside the tidal radius will likely become unbound (temporarily) at perigalacticon, where the tidal radius reaches its minimum. When the cluster moves away from perigalacticon and the instantaneous tidal radius of the cluster increases again, some stars are able to be re-captured by the cluster \citep{kupper2012}. Therefore, the Jacobi radius of a cluster will be greater than the perigalactic tidal radius, and  there are no clear relationship between limiting radii and perigalactic distance  (see e.g., \citealt{odenkirchen1997}).

Another approach is using the orbit-averaged tidal radius \citep{brosche1999}. This approach is already tested by several authors.  For example, \cite{kupper2010} and \cite{kupper2012} found that the time averaged mean tidal radius of the cluster and not the perigalactic tidal radius is better approximation to reproduce the structure of tidal tails.   Moreover, N-body simulations by \cite{madrid2012} found that the half-mass radius of a GC is more likely imposed at $R_A$ than $R_P$.

Alternatively, one can use the instantaneous Jacobi (tidal) radius which corresponds to the distance of $L_1/L_2$ Lagrange points, as if the satellite was on a circular orbit of radius equal to its current distance from the host. \cite{webb2013} explored the influence  of the orbital eccentricity on the tidal radius using the direct \Nbody experiments and showed that the limiting radius is not imposed at $R_P$, instead, it traces the "instantaneous tidal radius" of the cluster at any point in the orbit. It is shown that the assumption of the instantaneous Jacobi radius is good approximation for satellite galaxies orbiting around the main host galaxy (e.g., see figure 5 of \citealt{gajda2016}). Here in this work we follow this approach and will show that with this assumption the semi-analytical models follow the N-body simulations remarkably well.

Using the mass and galactocentric distance of the model clusters at each time
step, we calculate the instantaneous Jacobi radius of each model which is not constant.  The oscillation of $R_G(t)$ leads to the oscillation of $r_J$ and then the Jacobi radius is not constant and the instantaneous Jacobi radius increases and decreases
periodically along the orbit between apogalactic and perigalactic radii.   In order to check the validity of our assumptions on the instantaneous Jacobi radius made in the semi-analytical models, the evolution of different parameters in semi-analytical models  is compared to a realistic \Nbody models.

In order to put the criteria for determining whether a star is bound or unbound we invoke a distance cutoff such that the cluster-centric distance of stars must be greater than the cluster instantaneous tidal radius for it to be unbound. This criterion has been used frequently in \Nbody modeling of star clusters (e.g. \citealt{takahashi2012, haghi2015}). It has also been suggested that a star's velocity plays a role in determining the unbound stars (e.g. \citealt{kupper2010, kupper2012}). Given the different definition of bound stars, the evolution of bound stars displays different oscillating and staircase pattern, during the pericentre passage. These artificial behaviours that emerge due to the definition of bound for  clusters, illustrate that it is not possible to have a universal definition of the bound stars for clusters on eccentric orbits. \cite{webb2013} found that  this additional criteria only effected a small percentage of simulated stars and did not change any of the \Nbody results. Moreover, \cite{xu-cai2016} showed that the differences between the evolution of cluster for  different definition of bound are small.  Therefore, for both N-body models and semi-analytical calculations we only assumed the Jacobi radius as a distance cutoff to determine the star to be considered unbound. These assumptions are tested in Section \ref{Sec:3} (see Figs \ref{evolution1} to \ref{diff-ecc3}) where we show that the \Nbody results are followed by our semi-analytical models and hence, our definition of bound, i.e., being within the instantaneous Jacobi radius, is a good approximation.

\section{Summary}\label{Sec:4}

Although modeling a few low-density medium-sized GCs of the Milky Way with $N=10^5$ stars, over their entire lifetime is now possible by a direct \Nbody approach, modeling the majority of the Galactic GCs with a realistic density distribution and large number of stars  (i.e., $N\geq5\times10^5$) over a Hubble time is still beyond our computational ability.

Therefore, the faster codes such as \textsc{mocca} and \textsc{emacss} allow us to model the large number of Galactic GCs with realistic initial density, size and number of stars. But, both methods are limited for clusters moving on circular orbits.

Following the approach introduced in \textsc{emacss} code, we extend the model to include the dynamical evolution of single-mass star clusters in more realistic elliptical orbits. We calculated the evolutionary equations of \textsc{emacss} code so that the eccentric orbits can be now handled properly. Our approach addresses both the pre-collapse and post-collapse evolutionary phases of star clusters over their entire life-time.

We compared the evolution of single-mass star clusters using the semi-analytical approach with the outcome of \Nbody simulations and showed that the evolution of all parameters are in good agreement with each other. Our models include different initial sizes, number of stars and orbital eccentricities such that they cover both Roche-volume filling and under-filling systems with focusing merely on eccentric orbits.

The main advantage of our adopted approach is the runtime of the models (seconds to minutes) which is several orders of magnitude shorter than both Monte Carlo (hours to days) and \Nbody methods (days to months).

We intend to use our method to study the evolution of GC systems and model the observed properties of star cluster populations in a time-dependent galaxy potential.

\section*{Acknowledgements}
We would like to thank the anonymous referee for his/her useful comments and suggestions which improved the quality of this work. We thank Holger Baumgardt and Antonio Solima for helpful discussions. Pouria Khalaj is a team member of the StarFormMapper project which has received funding from the European Union's Horizon 2020 research and innovation programme under grant agreement No 687528.



\bsp \label{lastpage} \end{document}